\documentclass[12pt,psfig]{article}
\usepackage{amssymb}
\usepackage{amsmath}
\usepackage{epsfig}
\makeatletter
\def\@cite#1#2{{[{#1}]\if@tempswa\typeout
{IJCGA warning: optional citation argument
ignored: `#2'} \fi}}

\newcount\@tempcntc
\def\@citex[#1]#2{\if@filesw\immediate\write\@auxout{\string\citation{#2}}\fi
  \@tempcnta\z@\@tempcntb\m@ne\def\@citea{}\@cite{\@for\@citeb:=#2\do
    {\@ifundefined
       {b@\@citeb}{\@citeo\@tempcntb\m@ne\@citea\def\@citea{,}{\bf ?}\@warning
       {Citation `\@citeb' on page \thepage \space undefined}}%
    {\setbox\z@\hbox{\global\@tempcntc0\csname b@\@citeb\endcsname\relax}%
     \ifnum\@tempcntc=\z@ \@citeo\@tempcntb\m@ne
       \@citea\def\@citea{,}\hbox{\csname b@\@citeb\endcsname}%
     \else
      \advance\@tempcntb\@ne
      \ifnum\@tempcntb=\@tempcntc
      \else\advance\@tempcntb\m@ne\@citeo
      \@tempcnta\@tempcntc\@tempcntb\@tempcntc\fi\fi}}\@citeo}{#1}}
\def\@citeo{\ifnum\@tempcnta>\@tempcntb\else\@citea\def\@citea{,}%
  \ifnum\@tempcnta=\@tempcntb\the\@tempcnta\else
   {\advance\@tempcnta\@ne\ifnum\@tempcnta=\@tempcntb \else
\def\@citea{--}\fi
    \advance\@tempcnta\m@ne\the\@tempcnta\@citea\the\@tempcntb}\fi\fi}
\makeatother

\def\f#1#2{\frac{#1}{#2}}

\newcommand{\gsim}{\lower.7ex\hbox{$\;\stackrel{\textstyle>}{\sim}\;$}}
\newcommand{\lsim}{\lower.7ex\hbox{$\;\stackrel{\textstyle<}{\sim}\;$}}
\newcommand{\be}{\begin{equation}}
\newcommand{\ee}{\end{equation}}
\newcommand{\bea}{\begin{eqnarray}}
\newcommand{\eea}{\end{eqnarray}}
\def\dd{{\tilde{\nabla}}^2}

\def\fd{{\tilde{\nabla}}^4}

\usepackage{amssymb}

\def\baselinestretch{1}
\begin{document}
\catcode`@=11
\newtoks\@stequation
\def\subequations{\refstepcounter{equation}%
\edef\@savedequation{\the\c@equation}%
  \@stequation=\expandafter{\theequation}
  \edef\@savedtheequation{\the\@stequation}
  \edef\oldtheequation{\theequation}%
  \setcounter{equation}{0}%
  \def\theequation{\oldtheequation\alph{equation}}}
\def\endsubequations{\setcounter{equation}{\@savedequation}%
  \@stequation=\expandafter{\@savedtheequation}%
  \edef\theequation{\the\@stequation}\global\@ignoretrue

\noindent}
\catcode`@=12
\begin{titlepage}
\title{{\bf Consistent long distance modification of gravity from inverse powers of the curvature}}
\vskip2in

\author{
{\bf Ignacio Navarro$$\footnote{\baselineskip=16pt E-mail: {\tt
i.navarro@damtp.cam.ac.uk}}} $\;\;$and$\;\;$ {\bf Karel Van
Acoleyen$$\footnote{\baselineskip=16pt E-mail: {\tt
karel.van-acoleyen@durham.ac.uk}}}
\hspace{3cm}\\
 $$ {\small $^{*}$DAMTP, University of Cambridge, CB3 0WA Cambridge, UK}\\
{\small $^{\dagger}$IPPP, University of Durham, DH1 3LE Durham, UK}.
}

\date{}
\maketitle
\def\baselinestretch{1.15}

\begin{abstract}
\noindent

In this paper we study long distance modifications of gravity
obtained by considering actions that are singular in the limit of
vanishing curvature. In particular, we showed in a previous publication that models that include inverse
powers of curvature invariants that diverge for $r\rightarrow 0$
in the Schwarzschild geometry, recover an acceptable weak field
limit at short distances from sources. We study then the
linearisation of generic actions of the form ${\cal L}=F[R,P,Q]$
where $P=R_{\mu\nu}R^{\mu\nu}$ and
$Q=R_{\mu\nu\lambda\sigma}R^{\mu\nu\lambda\sigma}$. We show that
for the case in which $F[R,P,Q]=F[R,Q-4P]$, the theory is ghost
free. Assuming this is the case, in the models that can explain
the acceleration of the Universe without recourse to Dark Energy
there is still an extra scalar field in the spectrum besides the
massless spin two graviton. The mass of this extra excitation is
of the order of the Hubble scale in vacuum. We nevertheless
recover Einstein gravity at short distances because the mass of
this scalar field depends on the background in such a way that it
effectively decouples when one gets close to any source. Remarkably, for the
values of the parameters necessary to explain the cosmic
acceleration the induced modifications of gravity are suppressed at the Solar
System level but can be important for systems like a galaxy.
\end{abstract}

\vspace{3cm}

\vskip-23cm \rightline{} \rightline{DAMTP-2005-105} \rightline{DCPT/05/142}
\rightline{IPPP/05/71}

\end{titlepage}
\setcounter{footnote}{0} \setcounter{page}{1}

\baselineskip=20pt

\section{Introduction}

Theories that can accommodate a long distance modification of
gravity are interesting phenomenologically, but also from a purely
theoretical perspective. Phenomenologically, because although
Einstein gravity has passed many tests in the Solar System, the
issues of Dark Matter and Dark Energy might well have something to
do with a non-standard behaviour of gravity at large distances.
And from a theoretical standpoint any such theory is worth
consideration because it has proved very hard to find consistent,
generally covariant theories of this type. In this article we
report in what we believe represents a consistent, large distance
modification of gravity that, besides being able to explain the
acceleration of the Universe, can also have interesting
implications for the Dark Matter problem. These modifications of
gravity are generated by including inverse powers of the curvature
in the action, or, more generally, considering actions that are singular in
the limit of vanishing curvature. These models were first proposed
to explain the cosmic acceleration without the need for Dark
Energy \cite{Capozziello:2003tk}, but it was soon realised that
the simplest possibility ($\Delta {\cal L}\propto 1/R$) can give problems with the weak field limit \cite{Dolgov:2003px}. This can
be easily understood noticing that models whose Lagrangian is an
arbitrary function of the scalar curvature are generically equivalent
to scalar-tensor theories of gravity. In
the models that can explain the acceleration of the Universe by
introducing inverse powers of the scalar curvature the
corresponding scalar has a mass that is proportional to some positive power of the scalar curvature. One can expect then that in regions where the scalar curvature is very low this excitation generates a new force that would be in conflict with experiments. And although a rigorous quantitative computation of the predictions of these theories for Solar System measurements is still lacking in the literature it is not clear that they constitute a viable alternative to Dark Energy\footnote{For instance, one of the conseqences of this scalar field is that one would
  infer a different value of Newton's
  constant from Newton's potential and light bending.}.

But the situation is quite different if we consider inverse powers of other
invariants. We showed in a recent publication \cite{Navarro:2005gh}
that one can generically recover an acceptable weak field limit in
these models at short distances provided one includes inverse powers
of curvature invariants that do not vanish for the Schwarzschild
geometry. In particular, we considered the action
\be S=\int \!\!d^4x\sqrt{-g}\frac{1}{16\pi
  G_N}\left[R-\frac{\mu^{4n+2}}{(aR^2+bP+cQ)^n}\right]\,,\label{action}\ee with positive $n$ and where
\be P \equiv R_{\mu\nu}R^{\mu\nu}\,\;\; {\rm and}\;\;\; Q \equiv
R_{\mu\nu\lambda\rho}R^{\mu\nu\lambda\rho}. \ee Here $G_N$ is
Newton's constant and $\mu$ a parameter with dimensions of mass.
In this theory corrections to standard cosmology will only occur
at the present epoch if we take $\mu\sim H_0$ and it has been
shown that it has cosmologically interesting accelerating
solutions \cite{Carroll:2004de} that provide a good fit to the SN
data \cite{Mena:2005ta} without recurring to Dark Energy. But we
also showed in \cite{Navarro:2005gh} that beyond a source
dependent distance given by \be r_c^{6n+4} \equiv \frac{\left(G_N
M\right)^{2n+2}}{\mu^{4n+2}}, \ee where $M$ is the mass of the
source, gravity is modified. Moreover, we computed the first
correction to Newton's potential in an expansion in powers of
$r/r_c$ yielding \be V(r)\simeq -\left[1-\alpha
\left(\frac{r}{r_c}\right)^{6n+4} + {\cal
O}\left(\left(\frac{r}{r_c}\right)^{12n+8}\right)\right]\frac{GM}{r},
\ee where $\alpha\sim 10^{-2}$. It is remarkable that with the
values of the parameters necessary to explain the cosmic
acceleration, this characteristic distance puts the departure
from Newtonian gravity well under control at Solar System scales, but can
make it important for larger systems like a galaxy, hence the
potential implications for the Dark Matter problem. But the theory
under consideration involves up to fourth order derivatives of the
metric. The existence of these higher derivatives suggests that
one would always find ghosts in a linearised analysis. In fact, it
can be argued by considering the Cauchy problem and the gauge
symmetries and constraints of the theory (see for instance
\cite{Hindawi:1995cu}) that general actions involving arbitrary
powers of the Riemann tensor will contain at most eight degrees of
freedom\footnote{If we also included derivatives of the Riemann
tensor in the action even more degrees of freedom would show up
\cite{Hindawi:1995cu}.}: two in the usual massless graviton, one
in a scalar field and five in a ghost-like massive spin two excitation.
The problem with the ghosts is that one has to accept negative
energy states in the theory or lose unitarity
\cite{Hawking:2001yt}. In the former case the background in which
they appear is expected to be unstable at least towards the
generation of small scale inhomogeneities (positive and negative
energy particles) since that configuration would dominate the
phase space volume available to the system. Therefore the
consistency of these theories has been put into question
\cite{Nunez:2004ts,Chiba:2005nz}. In this paper we will show that
with a suitable choice of parameters ($b=-4c$), the theory obtained from the action (\ref{action}) is ghost-free. In
fact we will show that generic Lagrangians of the form ${\cal
L}=F[R,Q-4P]$ are ghost free. As expected, in this case we just
add a single scalar degree of freedom to the gravitational sector,
on top of the massless spin two graviton. For theories of the type
(\ref{action}) that can explain the acceleration of the Universe
without Dark Energy, the mass of this excitation is $\sim H$.
Nevertheless we still recover an acceptable Newtonian limit at short
distances from sources because the linearisation of the theory
breaks down in this domain, and the extra scalar degree of freedom
decouples.

At this point it is convenient to recall the properties of other
theories that offer long-distance modifications of gravity, since
we will see that all of them share common features that are
expected on general grounds for this kind of theories (see for
instance the discussion in \cite{Dvali:2004ph}). The simplest way
to modify gravity at large distances would seem to be to give the
graviton a tiny mass, $m_g$. This is not as simple as it seems,
because when doing so we also introduce more degrees of freedom in
the gravitational sector, and this has some unexpected
consequences. If we study the theory at a linearised level, the
only ghost-free mass term we can add is the Fierz-Pauli mass term \cite{Fierz:1939ix}.
When adding this mass term one finds out that the linearised
theory does not recover the results of Einstein gravity when we
take the massless limit $m_g \rightarrow 0$. This is the so-called
vDVZ discontinuity \cite{vanDam:1970vg}, and although it was first
interpreted as a failure of massive theories of gravity, it was
later interpreted as a failure of the linearisation
\cite{Vainshtein:1972sx,Deffayet:2001uk}. This is so because the new
extra degrees of freedom that the Fierz-Pauli term introduces have
couplings that are singular in the massless limit. This can be
qualitatively understood noticing that the ``kinetic term'' of the
extra degrees of freedom comes, roughly, from the  Fierz-Pauli
term. Once the fields are rescaled according to their canonical
normalisation, the mass of the graviton appears suppressing
non-renormalisable operators in the expanded action. For a small
graviton mass one finds that the ``cut-off'' of the theory is very
low and the linearisation breaks down when one gets close to a
mass. Therefore the results obtained from the linearised
Lagrangian are not valid at short distances in the spacetime of a
spherically symmetric mass. The distance at which the linearisation
breaks down is
usually called the Vainshtein radius since he was the first to
argue that in a fully non-linear generally covariant theory that
reduces to massive gravity upon linearisation one should be able
to check that the vDVZ discontinuity is absent
\cite{Vainshtein:1972sx}. Continuity in the massless graviton
limit should then be achieved non-perturbatively. Unfortunately it
is not possible to obtain this limit if we don't have a full
non-linear generally covariant completion of the Pauli-Fierz term,
and the issue of the consistency of massive gravity remains open.

The story is very similar in other more elaborate, generally covariant models that also yield a long-distance
modification of gravity. For instance models involving Lorentz violating
condensates for some extra fields have been
considered \cite{Arkani-Hamed:2003uy} to
achieve this kind of modifications, and these models have been shown to be
consistent. Also theories obtained from
braneworlds like the GRS \cite{Gregory:2000jc} and DGP \cite{Dvali:2000hr} models show this
kind of phenomenology. But while
the GRS model presents ghosts upon linearisation, these are not
present in the DGP case. This last model is probably the best
studied long distance modification of gravity so far. It is related to massive gravity,
since the graviton acquires a ``soft'' mass in its propagator. In this model the
strong coupling scale and the nonlinear effects that imply the absence
of the apparent vDVZ discontinuity have been studied in detail (see $e.g.$ \cite{Lue:2005ya} and references therein). The
situation can be described as a shielding of the would-be strongly
coupled modes by non-linear effects at short distances. In this case gravity
becomes higher dimensional at long distances.

As we have mentioned, the model we will deal with shares many
common features with these theories: looking at the linearised
action on de Sitter background there is an apparent discontinuity
in the would-be Einstein gravity limit ($\mu \rightarrow 0$) due
to an extra scalar degree of freedom that becomes massless and
would seem to make the model not realistic. That is again resolved
once one takes into account that this linearised version of the
theory is no longer valid in regions where the curvature is large.
The extra scalar degree of freedom gets $shielded$ and decouples
at short distances, affecting the dynamics only at long distances
where the background curvature is small. Our model, however, does
not produce a massive graviton: the spin two excitations of the
metric remain massless. In the next section we will briefly
summarise how and why do we recover the Schwarzschild solution at
short distances, $r \ll r_c$, making explicit the shielding
mechanism. In the third section we will discuss the linearised
action for generic Lagrangians of the form ${\cal L}=F[R,P,Q]$ and
we will show that when $F[R,P,Q]=F[R,Q-4P]$, the theory is ghost
free. In the fourth section we will clarify the range of validity
of the linearised action, having in mind theories of the type
(\ref{action}) that can explain the acceleration of the universe
without Dark Energy. We will see that the ``cut-off'', or strong coupling scale of the
linearisation in vacuum is given by $\Lambda_s \sim (M_p
H^3)^{1/4}$ and we will obtain the Vainshtein radius for an object
of mass $M$ as $r_V \sim (G_NM/H^3)^{1/4}$. This does not mean
that the theory loses predictive power at higher energies or
shorter distances, but one has to use non-perturbative methods or
a different kind of expansion to obtain the results. Section five
contains the conclusions. In the Appendix we offer the technical
details of our derivations: we give a gauge invariant
decomposition of the degrees of freedom contained in the metric as
well as the action expanded up to bilinear order in terms of these
excitations and their propagators.

\section{Short distance solution}

In this section we will briefly summarise the results we obtained
in \cite{Navarro:2005gh}. As we said we started with the action
(\ref{action}). To study the Newtonian limit we want to find
spherically symmetric solutions to the equations \be
G_{\mu\nu}+\mu^{4n+2}H_{\mu\nu}=0\,,\ee where $G_{\mu\nu}$ is the
usual Einstein tensor and $\mu^{4n+2}H_{\mu\nu}$ is the extra term
generated by the inverse powers of the invariants in the action.
The main point is that, when evaluated in the Schwarzschild
solution, the extra term that appears in the equations of motion
is of order \be \mu^{4n+2}H_{\mu}^{\nu(0)} \sim
(GM/r^3)(r/r_c)^{6n+4},\ee where $r_c$ was defined in the
introduction. This shows that also $G_{\mu\nu}$ tends to zero as
$r\rightarrow 0$ and we indeed approach the Schwarzschild geometry in
this limit\footnote{Remember
  that the Schwarzschild solution is the only spherically
  symmetric solution to the equation $G_{\mu\nu}=0$.}. At small distances,
$r\ll r_c$, we can then consider a small perturbation of the
Schwarzschild solution and solve at first order in the
perturbations, obtaining the correction to the Newtonian potential
presented before. This amounts to an expansion of the exact
solution in powers of $r/r_c$, and is therefore only valid in the
$r\ll r_c$ region. Notice that an expansion in powers of $r/r_c$
signals a breakdown of the perturbative series in powers of $G_N$.
This feature is expected in any theory yielding an infrared
modification of gravity \cite{Dvali:2004ph} and it is related to
the fact that the extra degrees of freedom introduced by the
modification are shielded at short distances. The reason of the
shielding behaviour is clear in this case: the Kretschmann scalar
$Q$, when evaluated in the Schwarzschild solution reads
$Q=\frac{48(G_NM)^2}{r^6}$, while both $P$ and $R$ vanish. So,
assuming $c$ is not zero in (\ref{action}), the term
$\mu^{4n+2}H_{\mu\nu}$ has a shielding prefactor for spherically
symmetric solutions that goes like $\sim (r/r_c)^{6n+4}$,
effectively switching off at small distances the modifications of
the solution induced by this term. So we see that gravity is
modified only at large distances in this theory for the same
reasons why cosmology is modified only at late times, when the
mean curvature of the universe falls down to a value $R\sim
\mu^2$.

However, although we have seen that this modification of gravity
carries with it a shielding mechanism for massive sources, we have
not seen what is it shielding us from. One important requirement
for any proposed modification of General Relativity is the
existence of stable phenomenologically viable cosmological
backgrounds that are free of ghosts and tachyons, or where, at
least, their effects can be reduced to unobservable levels locally. For
investigating this issue it is convenient to study the
linearisation of the theory over one such background. This is the
goal of the next section.

\section{Linearisation of modified gravity}

In this section we are going to study the linearisation of a
generic modified theory of gravity. In the case of actions of the
type (\ref{action}), this expansion can be thought of as
complementary to the expansion over a Schwarzschild background
that we considered previously, one being valid at short
distances and the other only at large distances. But as we said
let us consider here a generic action built with the scalars $R$,
$P$ and $Q$: \be S=\int \!\!d^4x\sqrt{-g}\frac{1}{16\pi
  G_N}F[R,P,Q]\,.\ee
The background over which we consider the linearisation should be
a solution to the equations of motion in vacuum. If we consider at
the moment constant curvature maximally symmetric spacetimes the
possible values of the curvature can be found as the solutions of
\be \left(2F_Q+3F_P\right)R^2+6F_RR-12F=0 \ee where
$F_Q=\partial_Q F$, etc...

To determine the stability of these solutions and the particle
content of the theory we should expand the action in these
spacetimes up to second order in the fluctuations. It can be seen
\cite{Hindawi:1995cu,Chiba:2005nz} that this expansion will be the
same as that obtained from \bea S=\int
\!\!d^4x\sqrt{-g}\frac{1}{16\pi
  G_N}\left[-\Lambda + \delta R
+\frac{1}{6m_0^2}R^2-\frac{1}{2m_2^2}C^{\mu\nu\lambda\sigma}C_{\mu\nu\lambda\sigma}\right]
\label{expand-action},\eea where $C_{\mu\nu\lambda\sigma}$ is the
Weyl tensor and we have defined \bea \Lambda &\equiv& \left<F-RF_R
+R^2\left(F_{RR}/2-F_P/4-F_Q/6\right)+R^3\left(F_{RP}/2+F_{RQ}/3\right)  \right.\nonumber \\
&&\left.+R^4\left(F_{PP}/8+F_{QQ}/18+F_{PQ}/6\right)\right>_0 \\
\delta &\equiv&
\left<F_R-RF_{RR}-R^2\left(F_{RP}+2F_{RQ}/3\right)
\right.\nonumber \\
& & \left.
-R^3\left(F_{PP}/4+F_{QQ}/9+F_{PQ}/3\right)\right>_0 \\
m_0^{-2}& \equiv&
\left<\left(3F_{RR}+2F_P+2F_Q\right)+R\left(3F_{RP}+2F_{RQ}\right)\right.\nonumber \\
& & \left.
+R^2\left(3F_{PP}/4+F_{QQ}/3+F_{PQ}\right)\right>_0\\
m_2^{-2} &\equiv & -\left<F_P+4F_Q\right>_0. \eea Here $<...>_0$
denotes the value of the corresponding quantity
on the background. So we see that the situation for the
perturbations here is the same as in conventional Einstein gravity
supplemented with curvature squared terms. It is well known
\cite{Hindawi:1995cu} that in this case the gravitational sector
contains a massive scalar excitation of mass $\sim m_0^2$ and a
ghost-like massive spin two field with mass $\sim m_2^2$, besides
the usual massless graviton. So the ghost will be absent and there will be no ghost-like unstabilities in the theory as long as $m_2^{-2}=0$. In particular it is evident that actions of the form
$F[R,P,Q]=F[R,Q-4P]$ are ghost free (see also \cite{Comelli:2005tn}). But this expansion is only
valid in constant curvature spacetimes. We would like to consider
expansions in other FRW backgrounds that are also solutions of the
equations of motion in vacuum, since for some phenomenologically
interesting models de Sitter space is unstable (the scalar is a
tachyon) and the late time solution corresponds to a power-law FRW
cosmology \cite{Carroll:2004de}. We show in the Appendix that the
absence of the ghost for actions of the form $F[R,P,Q]=F[R,Q-4P]$
also translates into these backgrounds. This is not surprising
since, locally (at scales much less than $H^{-1}$), there should
be no significant differences between these
spacetimes\footnote{One way to think about the absence of the
ghost is that its mass has been taken to infinity. Its absence is
then an ``ultraviolet statement'' that should not depend on the
details of the background.}. Moreover, we give in the Appendix the
decomposition of the metric and the linearised action in terms of
the physical gauge invariant excitations and their propagators. Using then the equation of motion for the extra scalar excitation it is straightforward to address the stability of the de Sitter background. This scalar will satisfy the Klein-Gordon equation with a mass given by (see eq.(\ref{SEOM}) of the Appendix)
\be
m_s^2\equiv -H^2\left(\f{25}{4}+16\f{C_1}{C_2}\right).\label{ms}
\ee
For
perturbations with zero spatial momentum we then have:
$\phi(t)=\phi_0e^{\pm i m_st}$. Taking into account the definitions, eqs.(\ref{SS},\ref{norm}), this results in a spacetime metric \be
  ds^2=-(1-\phi_0e^{-3Ht/2\pm i 
m_st})dt^2+e^{2Ht}(1-\phi_0e^{-3Ht/2\pm i
  m_st})(dx^2+dy^2+dz^2)\,.\ee From this we can infer that the
background de Sitter metric is stable for $m_s^2>-9H^2/4$,
consistent with the results of \cite{Faraoni:2005vk}, for $F(R)$ theories.

\section{The limits of linearisation}

We will turn now our attention back to the original action (\ref{action}). We have seen in the previous section that in the case $b=-4c$ the theory is ghost free. But looking at the excitations over the vacuum it would seem that even if there is no ghost, the theory under
consideration, containing an extra scalar degree of freedom with a
tiny mass,
would be ruled out by fifth-force experiments and Solar System
observations, as was concluded in \cite{Chiba:2005nz}. We will see that this is not in fact a problem, if one takes into account that the linearised action we are
using to extract these conclusions has a very limited range of validity. In
the expanded action higher order terms will be suppressed by inverse powers of the background
curvature. This means that this expansion of the theory will break
down when the curvature of the fluctuations is extremely small, of the
order of the background one. In regions of high curvature we can not trust the linearised action obtained from
(\ref{expand-action}). So for a spherically symmetric mass there is a radius
such that inside it we can no longer trust the results obtained from our expanded
action. This is analogous to the Vainshtein radius of massive
gravity. In the next subsection we will argue that the strong coupling scale in vacuum for theories of the type (\ref{action}) (with $\mu \sim H$), or the cut-off of its linearised version, is $\Lambda_s \sim \left(M_p H^3\right)^{1/4}$, where $M_p^2 = (16\pi G_N)^{-1}$. This strong coupling scale corresponds to the scale suppressing non-renormalisable operators for the canonically normalised excitations. Notice that this does not mean that the theory loses predictive power above this scale. We know the ultra-violet completion of the theory, but at energies above this scale we should use non-perturbative methods or a different kind of expansion. For spherically symmetric solutions this gives a Vainshtein radius given by $r_V \sim \left(G_N M/H^3\right)^{1/4}$. In subsection 4.2 we will explicitly check this result by computing the higher order corrections in spherically symmetric solutions, and we will find that higher order corrections do indeed become dominant at distances less than $r_V$.

\subsection{Higher order corrections: the strong coupling scale and the Vainshtein radius}

To obtain the strong coupling scale, we have to figure out the
scales of the coefficients that go in front of the higher order
non-renormalisable (marginal) operators, after canonical
normalisation of the fields. We will assume that our Lagrangian is
given by a function $F[R,Q-4P]$ and we are in de Sitter spacetime
for definiteness, with $R=12H^2$. It will be convenient to write
$F[R,Q-4P]=\tilde{F}[R,U]$, where $U\equiv 5R^2/6+Q-4P$ because
$U$ is zero on the background. We consider as usual the weak field
expansion in powers of $h_{\mu\nu}$ where
$g_{\mu\nu}=g^0_{\mu\nu}+h_{\mu\nu}$ and the background metric is
\be
ds^2=g^0_{\mu\nu}dx^{\mu}dx^{\nu}=-dt^2+a(t)^2(dx^2+dy^2+dz^2)\,\ee
with $a(t)\equiv e^{Ht}$. Retaining only the highest dimension
operators at every order (these will dominate at high
energies\footnote{Remember that by ``high energies/momenta'' we
mean
  here $q\gg H$.}), the expansion in $h_{\mu\nu}$ of $U$ roughly looks
like: \be
U\sim(\partial^2h)^2+\left((\partial^2h)^2h+\partial^2h\partial
h\partial
h\right)+\left((\partial^2h)^2h^2+\ldots\right)+\ldots\,.\ee We
have suppressed all the indices so we write $\partial$ to denote
any space or time derivative and $h$ for any $h_{\mu\nu}$\,. For
the expansion of the Ricci Scalar we have: \bea
R&\sim&12H^2+a(t)^{-3/2}\left((\dd+3H\partial_0+\f{15}{2}H^2)\phi+(3\partial_0^2-2\dd+3H\partial_0-\f{45}{4}H^2)\tau\right)\nonumber\\
&&+\left(h\partial^2h+\partial h\partial
h\right)+\left((\partial^2h)h^2+(\partial
h)^2h\right)+\ldots\,,\eea where $\nabla^2$ is the
three-dimensional Laplacian and $\dd \equiv a(t)^{-2}\nabla^2$. We
write the full first order term explicitly (using the gauge
invariant decomposition given in the Appendix), because the
scaling dimension of this term is not what it seems. At first
sight this term seems to go like $\partial^2h$ at high energies.
However, as we show in the Appendix the propagators are such that
when doing perturbative calculations, it is more convenient to
replace $\phi$ with $\tilde{\phi}$ (see eq.(\ref{phitilde}) in the
Appendix). This replacement diagonalises the propagators and
cancels out the $\partial^2\tau$ and the $\partial \tau$ terms in
the first order term. From eqs.(\ref{phitilde},\ref{phioperator})
we find now for high energies/momenta: \be R^{(1)}\approx
a(t)^{-3/2}\left(\dd \tilde{\phi} +(\ldots)H^2\tau\right)\,.\ee
Here we have considered $\tau$ and $\tilde{\phi}$ separately,
since, as we will see later, the scaling of the $\tilde{\phi}$
propagator is non-canonical.

Using the general forms of the expansions of $U$ and $R$ we can
now write the general form of the higher order terms in the
expansion of the action, again only retaining the highest
dimension operator at each order. This expansion will take the
form \bea S & \simeq & S^{(2)} + M_p^2\int d^4x \left(
\left<\tilde{F}_{UR}\right>_0(\partial^2h)^2R^{(1)}+\left<\tilde{F}_{UU}\right>_0(\partial^2h)^4\right.
\nonumber \\
&&\left.
+\left<\tilde{F}_{UUR}\right>_0(\partial^2h)^4R^{(1)}+\left<\tilde{F}_{UUU}\right>_0(\partial^2h)^6+\ldots\right),\eea
where $S^{(2)}$ is the action expanded up to second order in $h$
(see the Appendix). In order to estimate the strong coupling
scale, we still have to canonically normalise the fields. We show
in the Appendix that for high energies $\omega$ and high momenta
$\tilde{k}$, the propagators of the six gauge invariant modes
contained in $h_{\mu\nu}$, go like: \bea
<\chi_{ij}\chi_{ij}>_0&\sim&\f{1}{M_p^2(\omega^2-\tilde{k}^2)}\nonumber\,,\\
<w_iw_i>_0&\sim&\f{1}{M_p^2\tilde{k}^2}\nonumber\,,\\
<\tau\tau>_0&\sim&\f{1}{M_p^2(\omega^2-\tilde{k}^2)}\,,\\
<\tilde{\phi}\tilde{\phi}>_0&\sim&\f{H^2}{M_p^2\tilde{k}^4}\,\nonumber.
\eea One can read off now the canonical normalisation of the
excitations: $h=h^c/M_p\,,$ except for \be
\tilde{\phi}=\f{H}{M_p}\tilde{\phi}^c \,.\ee So in general, at
high energies/momenta, the contribution of $\tilde{\phi}$ in a
certain higher order term will be subleading with respect to the
contribution of the other excitations. However, as we have shown
explicitly, there is an exception for $R^{(1)} \sim
\f{H}{M_p}\dd\tilde{\phi^c}$.  Taking into account these canonical
normalisations, the expansion of the action now takes the form:
\bea S&\sim& S^{(2)} +\int d^4x \left(
\f{1}{M_pH^3}(\partial^2h^c)^2\dd\tilde{\phi}^c+\f{1}{M_p^2H^6}(\partial^2h^c)^4
+\f{1}{M_p^3H^7}(\partial^2h^c)^4\dd\tilde{\phi}^c\right.\nonumber\\&&\left.+\ldots+
\f{1}{M_p^{2m-2}H^{4m-2}}(\partial^2h^c)^{2m}+\f{1}{M_p^{2m-1}H^{4m-1}}(\partial^2h^c)^{2m}\dd\tilde{\phi}^c+\ldots
\right)\,,\label{highorderaction}\eea and we see that the strong
coupling scale, given by the lowest scale suppressing
non-renormalisable operators, is $\Lambda_s=(M_pH^3)^{1/4}$.

It is now straightforward to translate this strong coupling energy
scale to the Vainshtein radius, defined as the distance where the
perturbative expansion of the spherically symmetric solution,
corresponding to a central mass source, breaks down. In terms of
Feynman diagrams the perturbative calculation of this (classical)
solution amounts to the calculation of the tree level expectation
values of $\tau$ and $\phi$, induced by the mass source with
coupling \be S_m=\int d^4x
\f{M}{2a(t)^{3/2}}\delta^3(x)\phi(x)\approx \int d^4x
\f{1}{2a(t)^{3/2}} \f{M}{M_p}
\delta^3(x)\left(2\tau^c+H\tilde{\phi}^c\right)
\,,\label{masssource}\ee where the last equation follows from
(\ref{phioperator}) in the limit of short (physical) distances
$\tilde{r}\equiv r a(t)\ll 1/H$ from the source. The Feynman
diagrams for the first order expectation values of $\tau^c$ and
$\tilde{\phi}^c$ are given in fig.1a. The definition of the
canonical excitations is such that the only dimensionful quantity
entering in the short distance propagators is the energy/momentum
$q$; this gives a canonical dimension 1 for $\tau^c$ and a
canonical dimension 0 for $\tilde{\phi}^c$.  With this in mind one
can easily estimate the short distance behaviour coming from a
certain diagram, by taking into account that the source has a
dimensionless factor $M/M_p$ for the $\tau^c$-coupling and a
dimension one factor $HM/M_p$ for the $\tilde{\phi}^c$-coupling.
From the first order diagrams we then find \be {\tau^c}^{(1)} \sim
\f{M}{M_p}\f{1}{\tilde{r}}\,,\,\,\,\,\,\,\,\,\,\,\,\,\,\,\tilde{{\phi}^c}^{(1)}\sim\f{HM}{M_p}\tilde{r}\,.\label{exp1}
\ee 

\begin{figure}[h]
  \begin{center}
    \epsfig{file=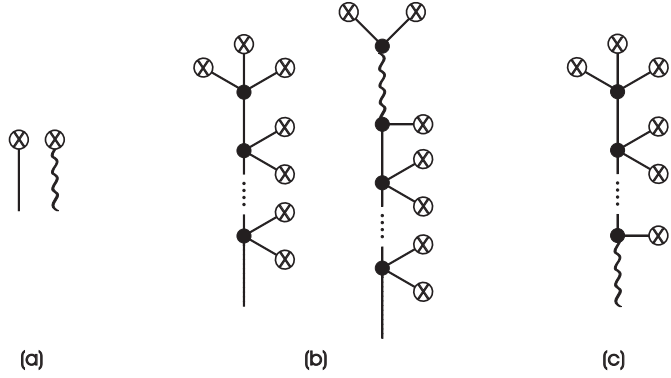,height=5cm}
  \end{center}
  \caption{\small Tree level tadpole Feynman diagrams for the expectation values of $\tau^c$ and $\tilde{\phi}^c$, induced by a mass
  source: the first order diagrams (a), some leading higher order
  diagrams for the expectation value of $\tau^c$ (b) and a leading higher
  order diagram for the expectation value of $\tilde{\phi}^c$ (c).  The full lines represent propagators of the fields $\chi_{ij},w_i$ and $\tau$. The $\tilde{\phi}$ propagator is represented as a wavy line.  \label{fig1}}
\end{figure}
The dominant higher order corrections will come from Feynman
diagrams containing only the third and fourth order vertex of
(\ref{highorderaction}), since these are the only ones suppressed
by $\Lambda_s$. We depict some of these higher order graphs in
fig.1b and fig.1c. Taking into account the relevant powers of
$\Lambda_s$ that come in front of the vertices, we now find that
the higher order ($m\geq1$) corrections go like: \bea
{\tau^c}^{(2m+1)}&\sim&
\left(\f{M}{M_p}\right)^{2m+1}\f{1}{\Lambda_s^{8m}}\f{1}{\tilde{r}^{8m+1}}\sim\f{M}{M_p}\f{1}{\tilde{r}}\left(\f{G_N
M}{H^3\tilde{r}^4}\right)^{2m}\,,\nonumber\\
\tilde{{\phi}^c}^{(2m)}&\sim&\left(\f{M}{M_p}\right)^{2m}\f{1}{\Lambda_s^{8m-4}}\f{1}{\tilde{r}^{8m-4}}\sim
\f{M^2}{M_p^3H^3}\f{1}{\tilde{r}^4}\left(\f{G_N
M}{H^3\tilde{r}^4}\right)^{2m-2}\,,\label{exp2}\eea and we see
that the expansions for $\tau^c$ and $\tilde{\phi}^c$, diverge for
distances $\tilde{r}$ smaller than \be r_V\equiv
\left(\f{M}{M_p}\right)^{1/4}\f{1}{\Lambda_s}\sim
\left(\f{G_NM}{H^3}\right)^{1/4}\,. \ee
Notice that we have not written down an expression for the even
orders in the $\tau^c$ expansion and of the odd orders in the
$\tilde{\phi}^c$ expansion. The reason is that the Feynman
diagrams corresponding to these orders unavoidably contain one or
more of the ``subleading'' higher order vertices in
(\ref{highorderaction}), suppressed by a mass scale bigger than
$\Lambda_s$. The coefficients of the even higher order terms for
the $\tau^c$ expansion, for instance, will then go as\be
{\tau^c}^{(2m)}\sim\f{M}{M_p}\f{1}{\tilde{r}}\left(\f{G_N
MH}{(H\tilde{r})^s}\right)^{2m-1}\label{exp3}\,,\ee where
$s\leq3$; and still show a convergent behaviour at distances
$\tilde{r}<r_V$, where the total series diverges.

\subsection{Explicit check: the Vainshtein radius}

In the previous subsection we derived diagrammatically, for
general Lagrangians of the form $F[R,Q-4P]$, the short distance
behaviour of the perturbative expansion of the spherically
symmetric solution corresponding to a central mass source. In
terms of the original fields $\tau$ and $\phi$, our results
(\ref{exp1}), (\ref{exp2}), (\ref{exp3}) for the expansions of
$\tilde{\phi}^c$ and $\tau^c$ become: \bea \tau&=&
X_1\f{G_NM}{\tilde{r}}+X_2\f{(G_NM)^2}{H^{s-1}\tilde{r}^{s+1}}+X_3\f{(G_NM)^3}{H^6\tilde{r}^9}+\ldots\,,\nonumber\\
\phi&=&
2X_1\f{G_NM}{\tilde{r}}+X_4\f{(G_NM)^2}{H^2\tilde{r}^{4}}+2X_3\f{(G_NM)^3}{H^6\tilde{r}^9}+\ldots\,.\eea
Here $X_i$ are dimensionless coefficients and we have used
$\phi=\tilde{\phi}+(2+\mathcal{O}(H^2\tilde{r}^2))\tau$.

Let us now consider a specific model, the action (\ref{action})
with $n=1$ (and $b=-4c$), and explicitly check that the first
orders in the expansions above do indeed exhibit this behaviour.
At linear order the equations of motion, obtained from (\ref{O1}),
(\ref{O2}), (\ref{O3}) and (\ref{masssource}) have the form: \bea
16\pi G_N\left(\frac{\delta
S}{\delta\tau}\right)^{(1)}=\fd\left(\f{C_2}{24H^2}(\phi-2\tau)\right)+\dd
\left((\f{C_2}{4H^2}\ddot{\tau}+\ldots)\right)+\ldots=0\label{EOM1}\\
16\pi G_N\left(\frac{\delta
S}{\delta\phi}\right)^{(1)}=-\f{1}{2}\fd\left(\f{C_2}{24H^2}(\phi-2\tau)\right)+
\dd\big(\ldots\big)+\ldots=-\f{8\pi
G_NM}{a(t)^{3/2}}\delta^3(x).\label{EOM2}\eea We will solve these
equations in a perturbative expansion in $\tilde{r}H$.  At short
distances, the dominant terms in the equations are the
$\fd$-terms. These terms originate from the term $\propto
{R^{(1)}}^2$ in the quadratic Lagrangian. At first sight, this
term would seem to give $\tau$ a propagator $\propto q^{-4}$.
However, as we discussed in the previous section, the propagator
has in fact the canonical $\propto q^{-2}$ behaviour. In the
equations of motion (\ref{EOM1},\ref{EOM2}) this is reflected in
the fact that the $\fd$-terms are proportional to each other,
which has consequences for the leading order (in $\tilde{r}H$)
behaviour of the solutions.  Looking at the 2nd equation one would
naively infer \be \tau,\phi\propto
a(t)^{3/2}G_NMH^2\tilde{r}(1+\mathcal{O}(\tilde{r}^2H^2))\,, \ee
since $\nabla^4r=-8\pi\delta^3(x)$. However, this is in conflict
with the first equation. The proper ansatze will be: \bea
\tau^{(1)}&=&a(t)^{3/2}\f{G_NM}{\tilde{r}}\big(\sum_{m=0}^{\infty}\f{d_m-c_m}{2}(\tilde{r}^2H^2)^m\big)\,\\
\phi^{(1)}&=&a(t)^{3/2}\f{G_NM}{\tilde{r}}\big(\sum_{m=0}^{\infty}(d_m+c_m)(\tilde{r}^2H^2)^m\big)\,.\eea
The coefficients are written in this peculiar form, because we
then have: $\phi-2\tau=\sum c_n(\ldots)$. Plugging in this ansatze
and solving the equations order by order we find: \bea
(\ldots)\nabla^4\f{1}{r}=0&\rightarrow& c_0=0 \,,\\
(\ldots)\delta^3(x)=0&\rightarrow&
d_0=\f{8}{3C_1}\,\,\,,c_1=-\f{3}{C_1}-\f{8}{C_2}\,,\\
(\ldots)\f{1}{r}=0&\rightarrow&d_1=-\f{c_1}{3}\,\,\,,c_2=\f{32}{3}\f{C_1}{C_2^2}+\f{4}{C_2}
\,.\eea The dimensionless constants $C_1$ and $C_2$ are defined as
\be C_1 \equiv \delta +\frac{4H^2}{m_0^2}\;\;\; {\rm and} \;\;\;
C_2 \equiv -\frac{16H^2}{m_0^2} \ee where $m_0$ and $\delta$ have
been defined in the third section. So we find for the spherically
symmetric solution, at first order in
$G_NM$:\bea\tau^{(1)}&=&a(t)^{3/2}\f{4}{3}
\f{G_NM}{C_1\tilde{r}}\big(1+(\f{3}{2}+\f{4C_1}{C_2})\tilde{r}^2H^2+\mathcal{O}(\tilde{r}^4H^4)\big)\\
\phi^{(1)}&=&a(t)^{3/2} \f{4}{3}
\f{G_NM}{C_1\tilde{r}}\big(2-(\f{3}{2}+\f{4C_1}{C_2})\tilde{r}^2H^2+\mathcal{O}(\tilde{r}^4H^4)\big)\,.\eea
Notice that the leading behaviour is precisely the one we expect
in scalar-tensor theories of gravity, but with a rescaled Planck mass $M_p^2 \rightarrow
C_1M_p^2$. We have also calculated the subleading correction
because this is required to obtain the leading correction for the
second order $\phi^{(2)}$ and $\tau^{(2)}$. These second order
corrections are the solutions to the equations: \bea
\left(\frac{\delta
S}{\delta\tau}\right)^{(1)}(\tau^{(2)},\phi^{(2)})&=&-\left(\frac{\delta
S}{\delta\tau}\right)^{(2)}(\tau^{(1)},\phi^{(1)})\,,\\
\left(\frac{\delta
S}{\delta\phi}\right)^{(1)}(\tau^{(2)},\phi^{(2)})&=&-\left(\frac{\delta
S}{\delta\phi}\right)^{(2)}(\tau^{(1)},\phi^{(1)})\,.\eea An explicit
calculation gives: \bea -16\pi G_N\left(\frac{\delta
S}{\delta\tau}\right)^{(2)}(\tau^{(1)},\phi^{(1)})=
-30\left(\f{4G_NM}{3C_1}\right)^2a(t)^{3/2}\f{12-12C_1+C_2}{4\tilde{r}^8H^4}(1+K_1\tilde{r}^2H^2+\ldots)\\
-16\pi G_N\left(\frac{\partial
S}{\partial\phi}\right)^{(2)}(\tau^{(1)},\phi^{(1)})=
15\left(\f{4G_NM}{3C_1}\right)^2a(t)^{3/2}\f{12-12C_1+C_2}{4\tilde{r}^8H^4}(1+K_2\tilde{r}^2H^2+\ldots)\,,\eea
with: \bea K_1&=&\f{732C_2+79C_2^2+96C_1-724C_1C_2-96C_1^2}{60C_2(12-12C_1+C_2)}\,\\
K_2&=&\f{240C_2+29C_2^2-240C_1-260C_2C_1+240C_1^2}{30C_2(12-12C_1+C_2)}\,.\eea
Notice that the leading terms in these equations have the same ratio as the $\fd$-terms in the first order equations of
motion. This can be traced back to the fact that they come from
the third order vertex in (\ref{highorderaction}). And now, in
contrast to the first order case, the leading behaviour will be
solution of $\nabla^4r^\alpha=1/r^8\rightarrow \alpha=4$. Writing then the
expansion as \bea
\tau^{(2)}&=&a(t)^{3/2}\left(\f{4G_NM}{3C_1}\right)^2\f{1}{\tilde{r}^4H^2}\left(\sum_{m=0}^{\infty}\f{f_m-e_m}{2}(\tilde{r}^2H^2)^m\right),\\
\phi^{(2)}&=&a(t)^{3/2}\left(\f{4G_NM}{3C_1}\right)^2\f{1}{\tilde{r}^4H^2}\left(\sum_{m=0}^{\infty}(f_m+e_m)(\tilde{r}^2H^2)^m\right)\,,\eea
and solving the equations order by order, we now find: \bea
(\ldots)\f{1}{r^8}=0&\rightarrow&
e_0=(12C_1-12-C_2)/(4C_2)\,\\
(\ldots)\f{1}{r^6}=0&\rightarrow&f_0=e_0\,\,,\,\,\,\,\,e_1=\ldots\,.\eea
So the second order terms in the perturbative solutions are: \bea
\tau^{(2)}&\propto&a(t)^{3/2}\f{(G_NM)^2}{\tilde{r}^2}\\
\phi^{(2)}&=&\f{12C_1-12-2C_1}{2C_2}a(t)^{3/2}\left(\f{4G_NM}{3C_1}\right)^2\f{1}{\tilde{r}^4H^2}(1+\mathcal{O}(\tilde{r}^2H^2))\,.\eea

 We will not obtain the actual 3rd order
correction, but just show that its leading behaviour is:
$\phi^{(3)}=2\tau^{(3)}\propto (G_NM)^3/(\tilde{r}^{9}H^6)$. We
find that the third order corrections are solutions of:\bea
\left(\frac{\delta
S}{\delta\tau}\right)^{(1)}(h^{(3)})&=&-\left(\frac{\delta
S}{\delta\tau}\right)^{(2)}(h^{(2)}\times
h^{(1)})-\left(\frac{\delta
S}{\delta\tau}\right)^{(3)}(h^{1})\\
&\approx& \f{a(t)^{3/2}}{16\pi G_N}\f{(4G_NM)^3}{(3C_1)^3
\tilde{r}^{11}H^6}\left(\f{12(8(C_1-1)+3C_2)(12+C_2-12C_1)^2}{C_2(C_2+4C_1-4)}\right)\,,\nonumber\\
\left(\frac{\delta
S}{\delta\phi}\right)^{(1)}(h^{(3)})&=&-\left(\frac{\partial
S}{\delta\phi}\right)^{(2)}(h^{(2)}\times
h^{(1)})-\left(\frac{\partial
S}{\delta\phi}\right)^{(3)}(h^{1})\\
&\approx& \f{a(t)^{3/2}}{16\pi
G_N}\f{(4G_NM)^3}{(3C_1)^3\tilde{r}^{11}H^6}\left(\f{6(16(1-C_1)-3C_2)(12+C_2-12C_1)^2}{C_2(C_2+4C_1-4)}\right)\,.\nonumber\eea
Now the leading terms in the right hand side of the equations of motion do not have
the special ratio that appeared in the
calculation of the second order correction. This comes without
surprise, since they originate from the fourth order vertex in
(\ref{highorderaction}) that does not contain an $R^{(1)}$ term.
This means that analogously to the first order case, the leading
behaviour of $\tau^{(3)},\phi^{(3)}$ will be solution of:
$\nabla^2 r^\alpha=r^{-11}\rightarrow \alpha=-9$ and we indeed find the
leading behaviour $\phi^{(3)}=2\tau^{(3)}\propto
(G_NM)^3/(\tilde{r}^{9}H^6)$. Notice that if we would have looked
only at the second order correction we would have inferred a wrong
Vainshtein radius ($\sim (G_NM/H^2)^{1/3}$). This was anticipated
by our feynmandiagrammatic derivation: the even orders in the
perturbation series for $\tau$ (and $\phi$) come with bigger
suppression scales than the strong coupling scale $\Lambda_s$, the
scale that determines the true Vainshtein radius.

\section{Conclusions}

Models that involve inverse powers of the Kretschmann scalar $Q$
in the action not only modify cosmology at late times, but also
Newtonian gravity at long distances. The reason behind this
behaviour is clear: in cosmology, when the Hubble scale is $H\gg
\mu$ the extra term that appears in the Friedmann equation is
suppressed by powers of $\mu/H$ and therefore negligible.
Analogously, for spherically symmetric solutions, the extra term
in the equations is now suppressed by powers of $r/r_c$, and this suppresses the modification at short distances where the usual
Einstein term dominates. Notice that the structure of the
spherically symmetric solutions for theories of the type
(\ref{action}) in a spacetime that has constant curvature
asymptotically is such that the scalar curvature goes to zero at
short distances and to a constant far from the source: this also
reflects the complete shielding of the modification of gravity at
short distances from sources that takes place in these theories.

We have also analysed the particle spectrum of the theory. In this
kind of theory, that involves up to fourth order derivatives of
the metric on its equations of motion, one can expect generically
eight propagating degrees of freedom in the gravitational sector:
two for the massless graviton, one for a scalar excitation and
five for a massive ghost. We have seen that the massive ghost is
absent as long as the action depends on $Q$ and $P$ only through
the combination $Q-4P$ \footnote{This is precisely the combination
appearing for instance in the Gauss-Bonnet term. It is amusing to
note that although the addition of this term to the action does
not affect the equations of motion in 4D, the addition of its
$inverse$ yields a ghost free long distance modification of
gravity.}. Moreover we have seen that the mass of the scalar field
depends on the background curvature. This enables us to offer a
physical picture in terms of particle excitations of the behaviour
of the solutions: the extra scalar field becomes massive, and
decouples, at short distances from sources. At long distances its
mass decreases to a very small value $m_S \sim H$ and its effects
can be noticed. So we are led to a situation in which gravity is
well described by conventional General Relativity at short
distances from sources, $r \ll r_c$, or in regions where the
curvature is much bigger than $\mu^2$. On the other hand, at
distances $r\gg r_V$, or in regions where the curvature is low,
gravity is well described by a scalar-tensor theory, obtained from
the linearisation of the action in vacuum. In both cases the force
of gravity will exhibit the characteristic $1/r^2$ falloff, albeit
with different proportionality constants. But there is still the
intermediate range, $r_c<r<r_V$, where there is a transition
between the two regimes, and where we can expect deviations from
the $1/r^2$ dependence in the force. To understand qualitatively
this transition we can consider the modification to the Newtonian
potential produced in scalar-tensor theories of gravity and argue
as follows. A massive scalar with mass $m_S$ and gravitational
couplings will typically contribute a term in the potential that
goes like \be V_S \sim \frac{G_NM}{r}e^{-m_Sr}, \ee with the
characteristic exponential Yukawa suppression of forces mediated
by massive fields. If the mass is of order $H$ we would expect
that the exponential will be well approximated by one everywhere.
But we have seen that in our case the situation is more
complicated because the mass of the scalar (and the Planck mass)
depends on the background curvature. As we have shown, in de
Sitter space the mass of the scalar is given by \be m_S^2 \sim
\frac{R_0^{2n+2}}{\mu^{4n+2}} \ee where $R_0$ is the background
curvature (or, more precisely, the vacuum expectation value of a
function of $R$, $P$ and $Q$). We expect then that, locally, the
mass of the scalar will have this dependence on the background
curvature. If we now take into account again that in spherically
symmetric solutions, a generic scalar involving $n$ powers of the
Kretschmann scalar is of order $Q^n \sim {\cal O} \left(
\left(GM\right)^{2n}/r^{6n}\right)$ at short distances, this
translates into an $r$-dependent contribution to the mass of the
scalar for actions of the type (\ref{action}) that goes like \be
\delta_{Source} m_S(r) \sim {\cal O}\left(
\left(GM/r^3\right)^{n+1}/\mu^{2n+1}\right).\ee Now we see that
the scalar mass has a contribution that grows at short radius. We
can obtain the distance $r_c$ as the distance at which the
``exponential factor'' in the contribution of the scalar to the
potential is of order one, $i.e.$ \be exp[-m_S(r)r] \sim 1\;\;\;
{\rm for} \;\;\; r \sim r_c.\ee For $smaller$ distances the would-be Yukawa
exponential factor suppresses the scalar contribution to the
potential, effectively decoupling this degree of freedom.

We have seen how this model shares many characteristic features
with other theories that also offer a long distance modification
of gravity, like massive gravity or the DGP model. In all of them
the perturbation theory valid at long distances from sources
breaks down at short distances. This characteristic shielding
behaviour makes it possible to avoid the vDVZ discontinuity, and
we recover an acceptable phenomenology even if there is an almost
massless scalar field in the spectrum. In the models that we have
studied here, however, we have a perfectly well defined theory
above the cut-off of the linearisation, in contrast with massive
gravity or other proposed long distance modifications of gravity.
One might still worry about the possible appearance of more
unwanted ghost-like degrees of freedom if one takes into account
the higher order terms in the expansion in powers of the
fluctuations of the gravitational equations, as happens in massive
gravity \cite{Boulware:1973my}. However, one can expect that the
mass of any such state, coming from the scale of suppression of
non-renormalisable operators, will be at least of the order of the
cut-off of the linearised theory, and therefore from an effective
theory point of view this would not really be a
problem\footnote{Similar remarks can be made regarding Einstein
  gravity. Quantum corrections generate higher order curvature terms
  in the action that inevitably produce ghosts upon linearisation. But the
mass of these apparent states is typically of the order of the Planck mass, so
they do not really represent a problem since one can not trust the
linearised action in this regime.}. As a
final cautionary remark we will also mention that although we have
proved the existence of acceptable solutions in the short
distance, $r\ll r_c$, and long distance $r\gg r_V$ regimes, we
have not proved that one can get a consistent matching between
them through the ``nonperturbative region'' (similar potential
problems have been pointed out for the case of massive gravity in
\cite{Damour:2002gp}). We regard this issue as more model
dependent, and we hope to address the behaviour of the solutions
in this intermediate regime more quantitatively in the future.

\section*{Acknowledgements}

We would like to thank D. Easson, G. Dvali, N. Kaloper, J.
Santiago and J. Weller for conversations and B. Solano for help
with the plot of the Feynman diagrams. K.V.A. is supported by a
postdoctoral grant of the Fund for Scientific Research - Flanders
(Belgium).

\section*{Appendix A. The gauge invariant degrees of freedom}

In this Appendix we will offer the decomposition of the fluctuations
of the metric in terms of gauge invariant combinations.
For the actions we will be interested in, Minkowski spacetime will
not be a solution so we have to expand on another metric. We will
take a cosmological FRW background metric: \be
ds^2=g^0_{\mu\nu}dx^{\mu}dx^{\nu}=-dt^2+a(t)^2(dx^2+dy^2+dz^2)\,.\ee
Writing $g_{\mu\nu}=g^0_{\mu\nu}+h_{\mu\nu}$ and leaving the
background metric ($g^0_{\mu\nu}$) fixed, an infinitesimal coordinate
transformation $x\rightarrow x+\xi$ gives: \be h'_{\mu
\nu}(x)=h_{\mu
\nu}(x)+\nabla_{\mu}\xi_{\nu}+\nabla_{\nu}\xi_{\mu}=h_{\mu
\nu}(x)+\nabla^0_{\mu}\xi_{\nu}+\nabla^0_{\nu}\xi_{\mu}+\mathcal{O}(h\xi)\,,\ee
where the superscript 0 indicates the covariant derivative on the
background metric. We will use the same decomposition as in
\cite{Rubakov:2004eb} to obtain the gauge invariant degrees of
freedom contained in the perturbations:
\bea h_{00}&=&\psi\nonumber\\
h_{0i}&=&u_i+\partial_iv\\
h_{ij}&=&\kappa_{ij}+(\partial_i s_j+\partial_j
s_i)+\partial_i\partial_j\sigma+ \delta_{ij}\rho\,. \eea Here
$\kappa_{ij}$ is a transverse-traceless tensor; $v_i$ and $s_i$
are transverse vectors, while the other fields are scalars under
global 3D rotations. At lowest order the tensor modes are gauge
invariant, while the other modes transform in the following way:
\bea \psi&\rightarrow& \psi
+ 2\partial_0\xi_0 \nonumber\\
u_i&\rightarrow& u_i+(\partial_0-2H)\xi^T_i\nonumber\\
v&\rightarrow& v+(\partial_0-2H)\eta+\xi_0\nonumber\\
s_i&\rightarrow& s_i +\xi^T_i\\
\sigma&\rightarrow&\sigma +2\eta\nonumber\\
\rho&\rightarrow&\rho-2a^2H\xi_0\,,\nonumber\eea where $\xi_i^T$ is
the transverse part of $\xi_i$, $\eta$ the longitudinal part
($\partial_i\xi_i=\nabla^2\eta$) and $H$ is the Hubble constant of
the background metric. There is one gauge-invariant combination in
the vector sector: \be w_i\equiv
u_i-(\partial_0-2H)s_i+\mathcal{O}(h^2)\,,\ee two in the scalar
sector: \bea
\phi&\equiv&\psi-2\partial_0v+\partial_0(\partial_0-2H)\sigma+\mathcal{O}(h^2)\,,\nonumber\\
\tau&\equiv&\rho+a^2H\left(2v-(\partial_0-2H)\sigma\right)
+\mathcal{O}(h^2)\,,\label{SS}\eea and finally there is one gauge-invariant
(transverse-traceless) tensor:
\be\chi_{ij}\equiv\kappa_{ij}+\mathcal{O}(h^2)\,.\ee Obviously,
these gauge invariant modes are the only ones that will show up in
the Lagrangian. Furthermore, the background is invariant under 3D
rotations, which implies that at the quadratic level a generic action
will have the following form: \be S^{(2)}=\frac{1}{16\pi
G_N}\int\!\!\!
d^4x\,\left(\chi_{ij}\hat{O}_\chi\chi_{ij}+w_i\hat{O}_w
w_i+\phi\hat{O}_{1}\tau+\tau\hat{O}_{2}\tau+\phi\hat{O}_3\phi\right)\,,\label{S2}\ee
where the operators $\hat{O}$ depend explicitly on time and
contain time derivatives and three-dimensional Laplacians.

\section*{Appendix B. The quadratic action}

\subsection*{B1. The case of Einstein gravity}

As a warming up exercise we will briefly review here the situation
in Einstein gravity for the gauge invariant decomposition we are
considering. In this case $F[R,P,Q]=R$ and the linearised action
in flat space one gets after integrating by parts and discarding
total derivatives is \be
 S^{(2)}=\frac{1}{64\pi G_N}\int\!\!\!
d^4x\;\left(-\chi_{ij}\Box^2\chi_{ij}-2w_i\nabla^2
w_i+4\phi\nabla^2\tau+6\tau\partial_t^2\tau-2\tau\nabla^2\tau\right).
\ee where $\Box^2$ is the d'Alembertian and $\nabla^2$ is the
three-dimensional Laplacian. At first sight the action doesn't
even look Lorentz invariant. But this is a constrained action and
the propagating modes do have Lorentz invariant dispersion
relations. The field $\phi$ does not represent any propagating
degree of freedom since it appears in the action as a Lagrange
multiplier. And it is clear from the equations of motion for $w_i$
and $\phi$ that the vector and the scalar $\tau$ also don't carry
any propagating degrees of freedom: \be \frac{\delta S^{(2)}}{\delta \phi}\propto \nabla^2 \tau=0 \ee \be
\frac{\delta S^{(2)}}{\delta w_i}\propto \nabla^2
w_i=0. \ee The only propagating modes are the two components of
the massless graviton contained in $\chi_{ij}$ that have a Lorentz
invariant dispersion relation. The solution to the equations of
motion for $\phi$ and $\tau$ coupled to sources do however generate
Newton's potential.

\subsection*{B2. The (dis)appearance of ghosts in the tensor/vector sector}

For Lagrangians of the form ${\cal L}=F[R,P,Q]$ we can expect up
to fourth order derivatives in the action. So focusing for
instance on the tensor modes we can expect, in the FRW background
we are considering:\bea \int\!\!\! d^4x\;
\chi_{ij}\hat{O}_\chi\chi_{ij} &=&\int\!\!\!
d^4x\;\Big(\ddot{\chi}_{ij}F^0_2(t)\ddot{\chi}_{ij}
+\dot{\chi}_{ij}\left(F_1^1(t)\nabla^2+F_1^0(t)\right)\dot{\chi}_{ij}\nonumber\\
&&+{\chi_{ij}}\left(F_0^2(t)\nabla^4+F_0^1(t)\nabla^2+F_0^0(t)\right){\chi_{ij}}\Big)\,,\eea
where a dot denotes time differentiation. If $F^0_2$ is different
from zero, the field $\chi_{ij}$ will contain 4 degrees of
freedom, grouped into 2 degenerate pairs. From general
considerations we know that the residues of the poles in the
propagator will then come with opposite sign
\cite{Hawking:2001yt}, so one of the pairs will be a ghost. For
$F^0_2(t)$ we find:\be
F^0_2(t)=\frac{1}{4a(t)}\left<F_P+4F_Q\right>_0. \ee So we find that
for actions of the form $F[R,P,Q]=F[R,Q-4P]$ there is no ghost in
general FRW backgrounds\footnote{Notice that when this is the case
we also eliminate two degrees of freedom in the vector sector and
one in the scalar sector that together form the five degrees of
freedom of the massive ghost.}. From now on we will consider that
the action is indeed a function of $P$ and $Q$ only through the
combination $Q-4P$. Also, we will restrict our analysis to de
Sitter space in what follows since in a generic FRW background the
results will be qualitatively similar, but the equations become
more involved. So we take $a(t)=e^{Ht}$ and we find useful to
define the constants: \be C_1 \equiv \delta
+\frac{4H^2}{m_0^2}\;\;\; {\rm and} \;\;\; C_2 \equiv
-\frac{16H^2}{m_0^2} \ee where $m_0$ and $\delta$ have been
defined in the third section. After introducing some normalisation
factors \footnote{When we write some of the excitations explicitly
in section 4, we include these normalisation factors.} :\bea
\chi_{ij}
&\rightarrow&a(t)^{1/2}\chi_{ij}\,,\nonumber\\
w_i&\rightarrow&a(t)^{-1/2}w_i\,,\nonumber\\
\phi&\rightarrow&a(t)^{-3/2}\phi\,\label{norm},\\
\tau&\rightarrow&a(t)^{1/2}\tau\,,\nonumber\eea and defining \be
\tilde{\nabla}^2\equiv\frac{\nabla^2}{a(t)^2}\,, \ee we find that the
operators in the tensor/vector sector of the quadratic action (\ref{S2}) read:
\bea
\hat{O}_{\chi}&=&\frac{C_1}{4}\left(-\partial_0^2+\dd+\frac{9}{4}H^2\right)\,,\\
\hat{O}_w&=&-\frac{C_1}{2}\dd\,.
\eea
So we find that at the quadratic level, the excitations in the
vector/tensor sector are the same as for ordinary gravity, but
with a rescaled Planck mass: $M_p^2\rightarrow C_1M_p^2$. As long
as $C_1>0$, the energy of the massless spin 2 graviton will be
positive, or, in other words: the vector/tensor sector will be
ghost free. We discuss the content of the scalar sector in the
next section.

\subsection*{B3. The propagating degree of freedom in the scalar sector}

In the scalar sector, the operators appearing in the quadratic
action are (again, assuming that $F[R,P,Q]=F[R,Q-4P]$, in a de Sitter background): \bea
\hat{O}_1&=&-\frac{C_2}{16}\frac{\dd}{H^2}\partial_0^2+\frac{C_2}{24}\frac{\fd}{H^2}+\frac{3C_2}{16H}\partial_0^3-\frac{3C_2}{16H}\dd\partial_0
-\frac{9}{32}C_2\partial_0^2+\left(C_1+\frac{51}{64}C_2\right)\dd\nonumber\\
&&-H\left(3C_1+\frac{75}{64}C_2\right)\partial_0+3H^2\left(\frac{3}{2}C_1+\frac{75}{128}C_2\right),\label{O1}\\
\hat{O}_2
&=&-\frac{3C_2}{32H^2}\partial_0^4+\frac{C_2}{8H^2}\partial_0\dd\partial_0-\frac{C_2}{24}\frac{\fd}{H^2}+\left(\frac{3}{2}C_1+\frac{51}{64}C_2\right)\partial_0^2-
\left(\frac{1}{2}C_1+\frac{3}{32}C_2\right)\dd\nonumber\\
&&-H^2\frac{9}{4}(\frac{3}{2}C_1+\frac{75}{128}C_2)\,,\label{O2}\\
\hat{O}_3&=&-\frac{C_2}{96}\frac{\fd}{H^2}+\frac{3C_2}{32}\partial_0^2-\frac{7C_2}{32}\dd-H^2\left(\frac{3}{2}C_1+\frac{75}{128}C_2\right)\label{O3}\,.\eea
At first sight the scalar sector looks like a mess. Naively one
would expect 3 propagating degrees of freedom. Also, as in the
case of Einstein gravity, the action does not look Lorentz
invariant. We will show explicitly now that there is in fact only
one degree of freedom, and its dispersion relation is, as it
should be, Lorentz invariant at high energies or short distances.

To get the true degree of freedom, the easiest thing to do is to
look at the equations of motion. We find that a suitable
combination of the equations of motion for $\phi$ and $\tau$
clearly exhibits a constraint of the theory: \be 0=\frac{\delta
S^{(2)}}{\delta \phi}+\hat{O}_4\left(\frac{\delta S^{(2)}}{\delta
\tau}+\hat{O}_5\frac{\delta S^{(2)}}{\delta
\phi}\right)=-\frac{3C_1}{16\pi G_N}H^2(\tau+\phi)\,,\label{EOMC1}
\ee where $\hat{O}_4$ and $\hat{O}_5$ are defined as:
\bea\hat{O}_4 &\equiv &
\frac{27}{2}\frac{H^4}{\fd}-3\frac{H^2}{\dd}+9\frac{H^3}{\fd}\partial_0\,,\\
\hat{O}_5&\equiv&\frac{1}{3}\frac{\dd}{H^2}+\frac{3}{2}+\frac{1}{H}\partial_0\,.\eea
Using this constraint in another combination of the equations of
motion, namely \be0=\frac{\delta S^{(2)}}{\delta
\tau}+\hat{O}_5\frac{\delta S^{(2)}}{\delta
\phi}=\f{1}{16\pi
G_N}\left(\hat{O}_6\tau+\hat{O}_7(\tau+\phi)\right)\,,\label{EOMC2}\ee
were now $\hat{O}_6$ and $\hat{O}_7$ are defined as: \bea
\hat{O}_6&\equiv&-\f{C_2}{48}\f{\fd}{H^4}\left(\partial_0^2-\dd-H^2\left(\f{25}{4}+16\f{C_1}{C_2}\right)\right)\,,\nonumber\\
\hat{O}_7&\equiv&-\f{C_2}{48}\f{\fd}{H^4}\left(H\partial_0+\f{1}{3}\dd+\f{5}{2}H^2\right)\,,\eea
one arrives at: \be
\left(\partial_0^2-\dd-H^2\left(\frac{25}{4}+16\frac{C_1}{C_2}\right)\right)\tau=0\,.\label{SEOM}
\ee So there is only one scalar degree of freedom, with mass
$m_S^2=-H^2\left(\frac{25}{4}+16\frac{C_1}{C_2}\right)$.

We will now determine the sign of the energy of the  propagating
mode. One way to do it would be by calculating the ``energy-momentum
tensor'' of the gravitons, $T_{\mu\nu}\equiv-(1/8\pi G_N)\hat{G}^{(2)}_{\mu\nu}$, and plug
in the solution. Here, $\hat{G}^{(2)}_{\mu\nu}$ stands for the second
order term in the expansion in powers of the fluctuations $h_{\mu\nu}$ of the full gravitational equations. We will follow a different strategy and look at the sign of the
residue of the pole in the propagators. In particular we will be looking at
the situation for large energy and 3-momenta: $\omega,\tilde{k}\equiv k/a(t)\gg H$, where the mode respects the Lorentz symmetry and the
propagators are diagonal in Fourier space. We will also discuss the
behaviour of the propagators for large off shell 4-momenta, since this information is useful for the discussion of the strong coupling scale in section 4.
After a little algebra we find the following expressions for the
propagators: \bea \hat{P}_{\tau\tau}&=&\left(\hat{O}_2-\frac{1}{4}\hat{O}_1^T\hat{O}_3^{-1}\hat{O}_1\right)^{-1}\,\nonumber\\
\hat{P}_{\phi\tau}&=&-\frac{\hat{O}_3^{-1}\hat{O}_1}{2}\hat{P}_{\tau\tau}\,\\
\hat{P}_{\phi\phi}&=&\left(\hat{O}_3-\f{1}{4}\hat{O}_1\hat{O}_2^{-1}\hat{O}_1^T\right)^{-1}\nonumber\\
&=&\hat{O}_3^{-1}+\frac{\hat{O}_3^{-1}\hat{O}_1}{2}\hat{P}_{\tau\tau}\frac{\hat{O}_1^T\hat{O}_3^{-1}}{2}\,.\eea
Here the propagators are defined in such a way that for instance:
\be <\phi(x)\tau(y)>_0=8\pi G_N i<x|\hat{P}_{\phi\tau}|y> \,,\ee
where $<\phi(x) \tau(y)>_0$ is defined as the tree level
expectation value of $\phi(x)\tau(y)$ one obtains from the path
integral. Again, at a first glance the propagators seem to contain
more than one degree of freedom. Naively, we would think that for
high energies, we can forget about the explicit time dependence
and ignore the fact that the operators do not commute
($[\partial_0,\dd]\neq0$). For the $\tau \tau$ propagator we then
expect to find in Fourier space: \be
P_{\tau\tau}(\omega,\tilde{k}^2)\approx
\frac{O_3(\omega,\tilde{k}^2)}{O_2(\omega,\tilde{k}^2)O_3(\omega,\tilde{k}^2)-\frac{1}{4}O_1(\omega,\tilde{k}^2)^TO_1(\omega,\tilde{k}^2)}
\approx\f{O_3(\omega,\tilde{k}^2)}{(\ldots)\omega^6+(\ldots)\omega^4+(\ldots)\omega^2+(\ldots)}
\,,\ee implying 3 propagating modes. In reality, the coefficients
in front of $\omega^6$ and $\omega^4$ cancel out, but to obtain
this second cancellation one has to take into account the time
dependence of the background reflected in the non-commutativity of
the operators. The easiest way to do this is by using some
identities that follow from equating the operators in front of
$\tau$ and $\phi$ at both sides of equations (\ref{EOMC1}) and
(\ref{EOMC2}). Using\bea \f{\delta S^{(2)}}{\delta
\tau}&=&\f{1}{16\pi G_N}\left(2\hat{O}_2\tau+\hat{O}_1^T\phi\right)\,,\\
\f{\delta S^{(2)}}{\delta \phi}&=&\f{1}{16\pi
G_N}\left(2\hat{O}_3\phi+\hat{O}_1\tau\right)\,,\eea we then
obtain from (\ref{EOMC2}): \bea
2\hat{O}_2+\hat{O}_5\hat{O}_1&=&\hat{O}_6+\hat{O}_7\,,\label{identity1}\\
\hat{O}_1^T+2\hat{O}_5\hat{O}_3&=&\hat{O}_7\,.\nonumber\eea  And
from (\ref{EOMC1}) we get: \bea
(1+\hat{O}_4\hat{O}_5)\hat{O}_1+2\hat{O}_4\hat{O}_2&=&-3H^2C_1\,,\\
2(1+\hat{O}_4\hat{O}_5)O_3+\hat{O}_4\hat{O}_1^T&=&-3H^2C_1\,.\nonumber\eea
Some manipulations of these equations learn us now that: \be
\hat{O}_2-\f{1}{4}\hat{O}_1^T\hat{O}_3^{-1}\hat{O}_1=-\f{3}{4}C_1H^2\left(\hat{O}_4^{-1}\hat{O}_3^{-1}\hat{O}_4\hat{O}_6\right)\,.\ee
So the $\tau\tau$ propagator can be written as: \be
\hat{P}_{\tau\tau}=-\f{4}{3C_1H^2}\hat{O}_6^{-1}\hat{O}_4^{-1}\hat{O}_3\hat{O}_4\approx-\f{2}{3C_1}\f{1}{\partial_0^2-\dd}\,,\ee
where the last equation holds for high energies/momenta
$\omega,\tilde{k}\gg H$. Just as in the case of the spin 2
graviton, we see that the scalar will have positive energy (or
positive norm), for $C_1>0$.

For feymandiagrammatical calculations, it is useful to project out
the propagating mode and work with $\tilde{\phi}$ instead of
$\phi$, where we define the former as \be \tilde{\phi} \equiv \phi
+\f{\hat{O}_3^{-1}O_1}{2}\tau \,. \label{phitilde}\ee The
$\tilde{\phi}\tilde{\phi}$ propagator then reads: \be
\hat{P}_{\tilde{\phi}\tilde{\phi}}=\hat{O}_3^{-1}\,.\ee and the
$\tilde{\phi}\tau$ propagator is zero. Notice, that for high
energies/momenta the $\tilde{\phi}\tilde{\phi}$ propagator has the
non-canonical behavior \be
\hat{P}_{\tilde{\phi}\tilde{\phi}}\approx -\f{96
H^2}{C_2}\f{1}{\fd}\,,\ee which is subleading with respect to the
canonical $\tau \tau$ propagator. When dealing with the issue of
the strong coupling scale in section 4 we need the explicit high
energy behaviour of $\hat{O}_3^{-1}\hat{O}_1$: \be
\hat{O}_3^{-1}\hat{O}_1 \approx
-2\left(2-3\tilde{\nabla}^{-2}\partial_0^2-9H\tilde{\nabla}^{-2}\partial_0
+9H\tilde{\nabla}^{-4}\partial_0^3\right)\,. \label{phioperator}
\ee Notice that, on shell ($\omega^2=\tilde{k}^2$), we have
$\hat{O}_3^{-1}\hat{O}_1\approx 2$, or $\phi=-\tau +\tilde{\phi}$,
consistent with the equation of motion $\tau+\phi=0$ for the
propagating mode.


\begin{thebibliography}{1}

\bibitem{Capozziello:2003tk}
S.~Capozziello, S.~Carloni and A.~Troisi,
arXiv:astro-ph/0303041;
S.~M.~Carroll, V.~Duvvuri, M.~Trodden and M.~S.~Turner,
Phys.\ Rev.\ D {\bf 70} (2004) 043528
[arXiv:astro-ph/0306438].

\bibitem{Dolgov:2003px}
A.~D.~Dolgov and M.~Kawasaki,
Phys.\ Lett.\ B {\bf 573} (2003) 1
[arXiv:astro-ph/0307285];
T.~Chiba,
Phys.\ Lett.\ B {\bf 575} (2003) 1
[arXiv:astro-ph/0307338].

\bibitem{Navarro:2005gh}
I.~Navarro and K.~Van Acoleyen,
Phys.\ Lett.\ B {\bf 622} (2005) 1
[arXiv:gr-qc/0506096].

\bibitem{Carroll:2004de}
S.~M.~Carroll, A.~De Felice, V.~Duvvuri, D.~A.~Easson, M.~Trodden and M.~S.~Turner,
Phys.\ Rev.\ D {\bf 71} (2005) 063513
[arXiv:astro-ph/0410031].

\bibitem{Mena:2005ta}
O.~Mena, J.~Santiago and J.~Weller,
arXiv:astro-ph/0510453.

\bibitem{Hindawi:1995cu}
A.~Hindawi, B.~A.~Ovrut and D.~Waldram,
Phys.\ Rev.\ D {\bf 53} (1996) 5597
[arXiv:hep-th/9509147].

\bibitem{Hawking:2001yt}
S.~W.~Hawking and T.~Hertog,
Phys.\ Rev.\ D {\bf 65} (2002) 103515
[arXiv:hep-th/0107088].

\bibitem{Nunez:2004ts}
A.~Nunez and S.~Solganik,
Phys.\ Lett.\ B {\bf 608} (2005) 189
[arXiv:hep-th/0411102].

\bibitem{Chiba:2005nz}
T.~Chiba,
JCAP {\bf 0503} (2005) 008
[arXiv:gr-qc/0502070].

\bibitem{Dvali:2004ph}
G.~Dvali,
arXiv:hep-th/0402130.

\bibitem{Fierz:1939ix}
M.~Fierz and W.~Pauli,
Proc.\ Roy.\ Soc.\ Lond.\ A {\bf 173} (1939) 211.

\bibitem{vanDam:1970vg}
H.~van Dam and M.~J.~G.~Veltman,
Nucl.\ Phys.\ B {\bf 22} (1970) 397,
V.~I. ~Zakharov, JETP Lett.\ {\bf 12}, 312 (1970).

\bibitem{Vainshtein:1972sx}
A.~I.~Vainshtein,
Phys.\ Lett.\ B {\bf 39} (1972) 393.

\bibitem{Deffayet:2001uk}
C.~Deffayet, G.~R.~Dvali, G.~Gabadadze and A.~I.~Vainshtein,
Phys.\ Rev.\ D {\bf 65} (2002) 044026
[arXiv:hep-th/0106001].

\bibitem{Arkani-Hamed:2003uy}
N.~Arkani-Hamed, H.~C.~Cheng, M.~A.~Luty and S.~Mukohyama,
JHEP {\bf 0405} (2004) 074
[arXiv:hep-th/0312099];
B.~M.~Gripaios,
JHEP {\bf 0410} (2004) 069
[arXiv:hep-th/0408127];
M.~V.~Libanov and V.~A.~Rubakov,
JHEP {\bf 0508} (2005) 001
[arXiv:hep-th/0505231].

\bibitem{Gregory:2000jc}
R.~Gregory, V.~A.~Rubakov and S.~M.~Sibiryakov,
Phys.\ Rev.\ Lett.\  {\bf 84} (2000) 5928
[arXiv:hep-th/0002072].

\bibitem{Dvali:2000hr}
G.~R.~Dvali, G.~Gabadadze and M.~Porrati,
Phys.\ Lett.\ B {\bf 485} (2000) 208
[arXiv:hep-th/0005016].

\bibitem{Lue:2005ya}
A.~Lue,
arXiv:astro-ph/0510068.

\bibitem{Comelli:2005tn}
D.~Comelli,
Phys.\ Rev.\ D {\bf 72} (2005) 064018
[arXiv:gr-qc/0505088].

\bibitem{Faraoni:2005vk}
V.~Faraoni and S.~Nadeau,
arXiv:gr-qc/0511094;
V.~Faraoni,
Phys.\ Rev.\ D {\bf 72} (2005) 061501
[arXiv:gr-qc/0509008].

\bibitem{Damour:2002gp}
T.~Damour, I.~I.~Kogan and A.~Papazoglou,
Phys.\ Rev.\ D {\bf 67} (2003) 064009
[arXiv:hep-th/0212155].

\bibitem{Boulware:1973my}
D.~G.~Boulware and S.~Deser,
Phys.\ Rev.\ D {\bf 6} (1972) 3368;
P.~Creminelli, A.~Nicolis, M.~Papucci and E.~Trincherini,
JHEP {\bf 0509} (2005) 003
[arXiv:hep-th/0505147].

\bibitem{Rubakov:2004eb}
V.~A.~Rubakov,
arXiv:hep-th/0407104.




\end{thebibliography}
\end{document}